\documentclass[twoside,twocolumn]{article}
\usepackage{blindtext} 
\usepackage[numbers]{natbib} 
\usepackage{amsmath}
\usepackage{amssymb}
\usepackage{graphicx}
\graphicspath{{images/}{../images/}}
\usepackage{caption, threeparttable}
\usepackage{afterpage}
\usepackage[sc]{mathpazo} 
\usepackage[T1]{fontenc} 
\usepackage{microtype}
\usepackage[english]{babel} 
\usepackage[hmarginratio=1:1,top=32mm,columnsep=20pt]{geometry} 
\usepackage{booktabs} 
\usepackage{lettrine} 
\usepackage{enumitem}
\setlist[itemize]{noitemsep} 
\usepackage{abstract}

\usepackage{titlesec} 
\renewcommand\thesection{\Roman{section}} 
\renewcommand\thesubsection{\roman{subsection}} 
\titleformat{\section}[block]{\large\scshape\centering}{\thesection.}{1em}{} 
\titleformat{\subsection}[block]{\large}{\thesubsection.}{1em}{} 
\usepackage{setspace}
\usepackage{fancyhdr} 
\usepackage{titling} 
\usepackage{hyperref} 

\pretitle{\begin{center}\Large\bfseries} 
\posttitle{\end{center}} 
\title{Comorbid CAD and Ventricular Hypertrophy Compromise The Perfusion of Myocardial Tissue at Subcritical Stenosis of Epicardial Coronaries} 
\author{%
\textsc{Eslam Abbas}\\[1ex] 
\normalsize Kobri El Kobba Medical Complex, Cairo, Egypt \\
}
\date{} 
\renewcommand{\maketitlehookd}{%
\begin{abstract}
\noindent \textbf{BACKGROUND:} Most studies of CAD revascularization have been based on and reported according to angiographic criteria which don't consider the relation between the resulting effective flow distal to the stenosis and the demand of a hypertrophied myocardial tissue.\\

\noindent
\textbf{MODEL:} Mathematical model of the myocardial perfusion in comorbid CAD and ventricular hypertrophy using Poiseuille's law. The analysis yields that the curve, which represents the relation between the perfusion and the severity of CAD depending on angiographic and/or angiophysiologic criteria, is shifted to the right by the effect of myocardial tissue hypertrophy. The right shift of said curve, which is directly proportional to the degree of ventricular hypertrophy, indicates that the perfusion of the corresponding myocardial tissue is compromised at angiographically and/or angiophysiologically subsignificant stenosis of the supplying epicardial vessel.\\

\noindent
\textbf{RESULTS:} Patients with comorbid CAD and left ventricular hypertrophy are more sensitive to CAD-related hemodynamic changes. They are more prone to develop ischemic complications, than their peers with isolated CAD regarding the same degree of coronary stenosis.\\

\noindent
\textbf{CONCLUSION:} Patients with comorbid CAD and ventricular hypertrophy suffer from myocardial hypoperfusion at angiographically and/or angiophysiologically subcritical epicardial stenosis. Accordingly; the comorbidity of both diseases should be considered upon designing of the treatment regime.

\end{abstract}
}

\begin{document}

\maketitle


\section*{Background}
Combined coronary artery disease and ventricular hypertrophy are not uncommon; they both share hypertension, which affects \(\sim 26\%\) of the world population \cite{kearney2005global}, as a risk factor. Accumulation of atheromatous plaques under tunica intima of the epicardial arteries restricts the blood flow to the supplied cardiac tissue. Chronic high-grade narrowing of the coronary arteries induces subendocardial ischemia during the escalation of the myocardial oxygen demand throughout exercise or stress \cite{libby2005pathophysiology}. The strained myocytes release mediators like adenosine and bradykinin \cite{safran2001cardioprotective}\cite{parratt1997bradykinin} which in addition to stimulating coronary vasodilatation, irritate the nerve endings leading to anginal pain \cite{crea1990role}.\\

The treatment strategy for treating CAD aims to improve survival and/or relieve symptoms \cite{wijns2010guidelines} including dyspnea and stable angina pectoris. Said strategy usually involves anti-anginal medications and/or PCI, or CABG in case of complex CAD and/or left main involvement, for achieving those aims. Trials show that revascularization by PCI or CABG is more effective than a strategy of medical therapy alone, in relieving symptoms like angina and dyspnea. Besides; it improves the quality of life by reducing the use of anti-angina drugs and increasing exercise capacity \cite{chamberlain1997coronary}\cite{time2001trial}\cite{weintraub2008effect}\cite{erne2007effects}. However; studies indicate that PCI, as an initial management strategy in patients with stable coronary artery disease, did not reduce the risk of complications as myocardial infarction or other major cardiovascular events when added to optimal medical therapy \cite{boden2007optimal}\cite{hueb2004medicine}\cite{stergiopoulos2012initial}.\\

The guidelines recommend that ad hoc PCI should not automatically be applied after angiography \cite{windecker2014authors}; and emphasize the usefulness of optimal medical treatment for selected patients, which can reduce angina and the risk of myocardial infarction and stroke substantially and prevent progression of atherosclerosis in the entire vasculature. Consequently; the clinical decisions for management of CAD generally need to balance adherence to guidelines against judgments based on specific patient, operator, social, economic, and cultural factors \cite{lamy2011medical}. Generally; PCI and medical therapy should be viewed as complementary, rather than opposing, strategies \cite{blumenthal2000medical}. Patients with stable coronary artery disease and functionally significant stenoses, benefit from the combination therapy of PCI plus optimal medical therapy by showing greater symptomatic improvement \cite{pursnani2012percutaneous} and decreased the need for urgent revascularization. However; in patients without ischemia, the outcome appeared to be favorable with the optimal medical therapy alone \cite{de2012fractional}.\\

For revascularization decisions and recommendations; said significant stenosis has been defined by most studies of CAD revascularization, which have been based on and reported according to angiographic criteria, as \(\geq70\%\) diameter narrowing, and/or \(\geq50\%\) for left main CAD \cite{levine20112011}. Besides; angiophysiological criterion, such as an assessment of fractional flow reserve (FFR), has been used in deciding when revascularization is indicated. Thus, for recommendations about revascularization, coronary stenosis with FFR \(\leq0.8\) is also considered to be significant \cite{pijls1996measurement}. The standard values provided by both methods, and so the revascularization decision, don't consider the relation between the resulting effective flow distal to the stenosis and the demand of a comorbid hypertrophied myocardial tissue.


\section*{Model}

Hagen-Poiseuille law, which is an analytical solution to the Navier-Stokes equation \cite{bruus2007theoretical}, states that; the flow rate \(Q\) through a coronary vessel is directly proportional to the pressure gradient \(\Delta P\) between the aortic root and the right atrium; and inversely proportional to the resistance \(R\) within the vessel. Wherein; the resistance \(R\) is inversely proportional to the radius \(\alpha\) of the vessel elevated to the fourth power, and is directly proportional to the blood viscosity \(\mu\) and the vessel length \(\Delta l\). So, by considering a circular cross-section of the vessel:

\begin{equation*}
Q=\Delta P \frac{\pi \alpha^4}{8 \mu \Delta l}
\end{equation*}

when

\begin{equation*}
R= \frac{8 \mu \Delta l}{\pi \alpha^4}
\end{equation*}

so;
\begin{equation*}
Q \propto \frac{1}{R} \propto \alpha^4 \propto \frac{1}{\Delta l}
\end{equation*}\\

The blood flow, which is a non-Newtonian fluid, within the circulation doesn't imitate precisely this law \cite{klabunde2011cardiovascular}, because said relation is applied on a Newtonian fluid in the steady laminar flow moving through a long cylindrical pipe.  Still, the law outlines the dominant determinants which influence the blood flow \(Q\) within the vasculature either in physiological or pathological conditions.\\

Atherosclerosis commonly affects the epicardial coronary vessels leading to narrowing of the vessel caliber \(\alpha_e\) and increase vascular resistance of the supplying vessel \(R_e\).
While:

\begin{equation*}
R_e \propto \frac{1}{\alpha^4_e}
\end{equation*}

\noindent the corresponding supplied myocardial segment doesn't actually suffer this severe blood flow reduction indicated in the above equation. The vasculature of the coronary circulation is arranged in-series, in addition to the in-parallel arrangement, so that the epicardial vascular resistance \(R_e\) is a segmental resistance. The coronary circulation can be divided into two compartments, the large epicardial conduit vessels and the resistance vessels, which are typically less than \(300 \mu m\) in diameter \cite{schelbert2010anatomy}. Whereas the conduit vessels exert little if any resistance to flow, resistance to flow progressively rises as the vessel diameter of the resistance vessels declines from about \(300 \mu m\) in the small arteries to less than \(100 \mu m\) in the arteriolar vessels \cite{muller1996integrated}. Therefore; the total resistance to blood flow \(R\) comprises mainly the pre-capillary resistance \(R_c\), the resistance of microvasculature \(R_m\), and the negligible resistance of the epicardial or conductance vessels \(R_e\).

\begin{equation*}
R = R_c + R_m + R_e
\end{equation*}\\

Narrowing of the radius of the epicardial vessel, due to atheromatous plaque, will increase the resistance in this vessel; but as:

\begin{equation*}
R_c + R_m \gg R_e
\end{equation*}

\noindent the impact of mild to moderate increase of the epicardial resistance \(R_e\) on the overall resistance of the coronary circulation \(R\) is insignificant.\\

However; in case of combined coronary artery disease and ventricular hypertrophy, both \(R_e\) and \(R_c+R_m\) are increased. Microangiogenesis is activated during the pathogenesis of ventricular hypertrophy as a compensatory mechanism to maintain effective blood supply to the hypertrophied tissue. Accordingly; CAD causes an increase in \(R_e\) due to epicardial arterial stenosis and ventricular hypertrophy increases \(R_c+R_m\) due to neomicroangiogenesis, i.e. addition of a new microvascular segment.

\begin{equation*}
R_{c+m} \propto \frac{1}{\alpha^4_m} \propto \Delta l
\end{equation*}

\noindent consequently; the flow rate \(Q\), and so the perfusion of myocardial tissue, diminish significantly upon subcritical stenosis of the supplying epicardial artery during the pathogenesis of CAD.\\

As mentioned; the identification of clinically-significant stenosis of an epicardial artery depends on an angiographic criterion, its radius \(\alpha_e\), and/or an angiophysiologic criterion, the FFR, which is an absolute number result from the ratio of the pressure distal to the lesion, to the pressure proximal to the lesion during induced hyperemia \cite{pijls1996measurement}. The graphical representation of the relation between both these criteria of the supplying artery and the perfusion of the supplied myocardial tissue follows a direct proportional relationship represented by a sigmoid-shaped curve, due to the effect of segmental resistance. Myocardial perfusion \(\xi\) describes the blood flow \(Q\) in ml/min per cubic centimeter of cardiac muscle volume \(V\).

\begin{equation*}
\xi = \frac{Q}{V}
\end{equation*}\\

According to the relation between the radius \(\alpha_e\) of an epicardial coronary artery, as an angiographic criterion, and the perfusion \(\xi\) of the corresponding supplied myocardial tissue represented in Figure 1; the perfusion \(\xi\) doesn't decrease significantly with gradual stenosis in isolated CAD until a critical stenotic value \(\phi_\alpha\) is reached, wherein the perfusion collapses relatively. Clinically; said critical value \(\phi_\alpha\) is defined as  \(\geq70\%\) radius  \(\alpha_e\)  reduction, significant stenosis \cite{levine20112011}. However; in patients with comorbid CAD and ventricular hypertrophy; the curve is shifted to the right indicating an increase in the critical stenotic value \(\phi_\alpha\), so that the perfusion \(\xi\) of the corresponding supplied myocardial tissue collapses relatively at a clinically subsignificant stenosis. The right shift in said patients depends on the degree of ventricular hypertrophy.\\

Additionally; the relation between another angiophysiologic criterion, the fractional flow reserve FFR, within a stenotic epicardial artery and the perfusion \(\xi\) of the corresponding supplied myocardial tissue, as represented in Figure 2, indicates that the perfusion \(\xi\) is not meaningfully reduced with the gradual decrease of FFR until a critical value \(\phi_{FFR}\) is reached, wherein the perfusion \(\xi\) collapses relatively. Clinical trials defined said critical value \(\phi_{FFR}\) as a FFR \(\leq 0.8\) \cite{tonino2009fractional}\cite{pijls2007percutaneous}. Though; in patients with combined CAD and ventricular hypertrophy; the curve shows a right shift, which is directly proportional to the degree of ventricular hypertrophy, indicating an increase in the critical stenotic value \(\phi_{FFR}\), so that the perfusion \(\xi\) of the corresponding supplied myocardial tissue collapses relatively at a clinically subsignificant reduction in the FFR.\\

The proposed model gives a more sensitive formula to detect the critical stenosis, which takes into account the demand of the supplied bulky myocardium. The isolated CAD curve is a logistic function; wherein \(x\) represents the critical stenosis and \(k\) is the curve slope:

\begin{equation*}
f(x) = \frac{1}{1+e^{-kx}}
\end{equation*}

\noindent in patients with comorbid CAD and ventricular hypertrophy; the curve is shifted to the right by \(a\) yielding \(x^\backprime\) as a representation of the critical stenosis:

\begin{equation*}
f(x) = \frac{1}{1+e^{-k(x^\backprime - a)}}
\end{equation*}

\noindent then;

\begin{equation*}
x^\backprime = x+a
\end{equation*}

\noindent wherein the curve shift \(a\) is directly proportional to the difference in muscle bulk \(\Delta M\) which obtained by Echocardiogram;

\begin{equation*}
a \propto \Delta M
\end{equation*}

\begin{equation*}
a = \omega \Delta M
\end{equation*}

\noindent the value of the constant \(\omega\) can be obtained experimentally. So; the percentage of the critical patency in patients with comorbid CAD and ventricular hypertrophy \(x^\backprime\) is:

\begin{equation*}
x^\backprime = x + (\omega \Delta M)
\end{equation*}

\section*{Results}

\textbf{Individuals with pathological ventricular hypertrophy are more sensitive to haemodynamic changes of the coronary circulation or pathologies that reduce the coronary reserve.}  Ventricular hypertrophy stresses the subendocardial myotissue due to increase the structural resistance of the coronary circulation. Said stress is ameliorated by compensatory functional changes to sustain the normal coronary blood flow. Although during vigorous exercise, the compensatory capability of the coronary flow reserve is exhausted under the effect of demand upsurge and shortened diastolic period. Occasional hemodynamic disturbances or subclinical pathologies which lessen the maximum coronary reserve may lead to selective subendocardial hypoperfusion.\\

\textbf{Comorbid CAD and ventricular hypertrophy cause the subendocardial tissue to suffer, during exercise or stress, from ischemia at an angiographically subsignificant stenosis in the supplying epicardial artery.} CAD primes the structural resistance of the neomicrovasculature of the hypertrophied tissue.  So; subcritical stenosis of the corresponding epicardial artery, mainly due to atherosclerosis, causes the total resistance to rise effectively to reduce the flow rate and exhaust the reactive compensatory mechanisms. The curve shift to the right in said patients doesn't affect the risk of myocardial infarction, \textbf{yet they are more susceptible to and usually presented by NSTEMI; with higher rates of transition from ischemia to necrosis in the affected hypertrophied endocardial tissue.} Increased muscle bulk shifts the endocardium away from the main blood supply. Besides; subjection to higher extravascular pressure depletes the functional vasodilator reserve in long standing pathological hypertrophy.\\

\textbf{Patients with combined CAD and ventricular hypertrophy have a higher risk to develop arrhythmias than their peers who suffer from isolated CAD.} In pathological hypertrophy; the neomicroangiogenesis shows anatomical and architectural dysgenesis in relation to the hypertrophied tissue. Said dysgenesis leads to failure of the coronary bed to uniformly supply the cardiac muscle, rendering foci within the hypertrophied muscle bulk at greater risk of ischemic injury. These stressed foci can be arrhythmogenic upon increased cardiac demand leading to serious arrhythmia and sudden cardiac death.

\section*{Discussion}

During cardiac catheterization; the main determinants of revascularization therapy in CAD patients, are either angiographic or angiophysiological criteria, measured during drug induced hyperemia, to identify the clinically-significant stenosis. Said determinants depend on the relation between the size of the insinuated plaque and the vascular diameter. A stenosis which reduces the radius of the epicardial vessel by \(70\%\) is considered significant angiographically. In another determinant; a small sensor on the tip of the guidewire is used to measure the pressure, temperature and flow to determine the exact severity of the lesion. The ratio between the pressure distal to the lesion and the pressure proximal to the lesion, which is the fraction flow reserve FFR, measures the pressure drop caused by the stenosis. A fraction flow reserve value \(\leq 0.8\) defines the stenosis to be significant. The FFR is a better method to detect the physiological significance of a stenosis \cite{pijls2010fractional}, as it takes into account collateral flow, which can render an anatomical blockage functionally unimportant. Also, standard angiography can underestimate or overestimate narrowing, because it only visualizes contrast inside a vessel. Yet; the standard values provided by both methods to identify a stenotic lesion as significant, don't consider the relation between the resulting effective flow distal to the stenosis and the demand of a hypertrophied myocardial tissue.\\

The pathogenesis of ventricular hypertrophy implicates an increase in the number of force-generating sarcomeres in the myocyte \cite{lorell2000left}. According to La Place law; an increase in pressure can be offset by an increase in wall thickness \cite{gunther1979determinants}. This mechanical input is transduced into a biochemical sequel that modifies gene transcription in the nucleus. The focal adhesion complex, in addition to the G-coupled neurohormonal augmentation \cite{sugden1999signaling}, is proposed as the effector transducers of said mechanical input \cite{borg1992holding}. Integrins connect the internal cytoskeleton of the cell to the extracellular matrix, wherein multiple tyrosine-phosphorylated kinases and serine-threonine kinases that are implicated in the signaling of hypertrophy can be found in the ECM \cite{kuppuswamy1997association}\cite{terracio1991expression}.\\

Angiogenesis is triggered during the pathogenesis of myocardial hypertrophy by increased cardiac work and oxygen demand; in an attempt to normalize maximal myocardial perfusion and capillary domains to sustain oxygen delivery. A limitation of capillary growth will increase diffusion distance for oxygen, while inadequate arteriolar growth will reduce maximal tissue perfusion. Pathogenesis of hypertrophy is categorized into pressure overload-induced, volume overload-induced, thyroxin-induced and exercise-induced models according to the stimulus for increasing muscle bulk. In exercise-induced and thyroxin-induced models; angiogenesis and arteriogenesis are well documented experimentally \cite{tomanek2007angiogenesis}. While in other models; there is a considerable variation in the reports of the literature about the extent and pattern of angiogenesis and the consequential coronary microvascular resistance. The reasons for the discrepancy between these studies are not evident, but the duration of the hypertrophy and the specificity of the stimulus may play a role.\\

Mathematically; angiogenesis increases the coronary microvascular resistance structurally, due to the addition of a new microvascular segment. However, in vivo; structural resistance can be modulated by functional changes, wherein autoregulatory adjustments involving the vasodilator reserve may ameliorate said structural resistance escalation. Well-trained athletes with physiological cardiac hypertrophy show a proportional increase of cardiac myocytes and coronary vasculature with no change in the proportion of extracellular collagen \cite{duncker2008regulation}. These structural modulations are accompanied by functional adaptations resulting in a compensatory exponential coronary reserve and vasodilator capacity. Functional adaptations can include changes in neurohumoral control and changes in local vascular control mechanisms \cite{laughlin1985effects}\cite{laughlin2004physical}. In pathological severe hypertrophy; pathological features of the strained neoangiogenesis halt the functional compensation for the structural increase in the microvascular resistance. Endothelium-dependent vasodilation is markedly impaired in the coronary microvessels of patients with hypertension-induced ventricular hypertrophy \cite{treasure1993hypertension}. Accordingly; severe ventricular hypertrophy is associated with a reduction in coronary vascular reserve \cite{pichard1981coronary}.\\

Myocardial infarction is mainly caused by rupture of vulnerable fibroatheromatous plaque forming a thrombus that interferes with myocardial blood supply leading to excessive ischemia then necrosis \cite{reed2017acute}. Usually; soft non-stenotic plaques are more susceptible to rupture, causing major cardiovascular events \cite{libby2013mechanisms}. The vulnerability of the plaque depends on lesion-specific characteristics like thin fibrous cap, large lipid-rich necrotic core, increased plaque inflammation, positive vascular remodeling, increased vasa-vasorum neovascularization, and intra-plaque hemorrhage \cite{moreno2010vulnerable}. Therefore; the comorbidity between CAD and ventricular hypertrophy doesn't affect the risk of developing MI. However; patients with said comorbid diseases have higher rates of transition from ischemia to necrosis in the affected endocardial tissue, due to increase diffusion distance for oxygen and exhaustion of the functional compensation. They are also more susceptible to and usually presented by NSTEMI, due to the sensitivity of the endocardial myotissue to the equilibrium between the structural resistance of microvasculature and the reactive functional modulation.\\

Myocardial ischemia is characterized by ionic and biochemical alterations, creating an unstable electrical substrate capable of initiating and sustaining arrhythmias \cite{ghuran2001ischaemic}. Theoretically; onerous angiogenesis in pathological hypertrophy shows patterns of anatomical and architectural dysgenesis rendering foci within the hypertrophied muscle bulk at greater risk of ischemic injury. Said stressed foci acquire different electrochemical properties, due to defective function of ATPase-dependent pumps, leading to tissue heterogeneity. Theses foci become arrhythmogenic, especially with increased cardiac demand during above-normal exercise or severe stressful conditions, leading to functional re-entry. Hence; presence of ventricular hypertrophy is associated with a greater risk of sustained arrhythmias \cite{chatterjee2014meta}.
\section*{Conclusion}
The mathematical model establishes that ventricular hypertrophy increases the vascular structural resistance of the coronary circulation due to neomicroangiogenesis. Patients with comorbid CAD and ventricular hypertrophy suffer, due to exhaustion of functional compensation, from complications of myocardial hypoperfusion at angiographically sub-significant coronary artery stenosis. These patients are more susceptible to NSTEMI, serious arrhythmias and sudden cardiac death than patients with isolated CAD. Upon confirmation of such results by large investigational studies, said results should be taken into account during designing the treatment strategy of said patients.

\section*{Acknowledgement}
The author states that there is no conflict of interest regarding this article.
\thispagestyle{plain}
\bibliographystyle{unsrt}
\bibliography{myref}

\begin{thebibliography}{10}

\bibitem{kearney2005global}
Patricia~M Kearney, Megan Whelton, Kristi Reynolds, Paul Muntner, Paul~K
  Whelton, and Jiang He.
\newblock Global burden of hypertension: analysis of worldwide data.
\newblock {\em The lancet}, 365(9455):217--223, 2005.

\bibitem{libby2005pathophysiology}
Peter Libby and Pierre Theroux.
\newblock Pathophysiology of coronary artery disease.
\newblock {\em Circulation}, 111(25):3481--3488, 2005.

\bibitem{safran2001cardioprotective}
Noam Safran, Vladimir Shneyvays, Nissim Balas, Kenneth~A Jacobson, Hermann
  Nawrath, and Asher Shainberg.
\newblock Cardioprotective effects of adenosine a1 and a 3 receptor activation
  during hypoxia in isolated rat cardiac myocytes.
\newblock {\em Molecular and cellular biochemistry}, 217(1):143--152, 2001.

\bibitem{parratt1997bradykinin}
James~R Parratt, Agnes Vegh, I~Jack Zeitlin, Maqsood Ahmad, Keith Oldroyd,
  Karoly Kaszala, and Julius~Gy Papp.
\newblock Bradykinin and endothelial-cardiac myocyte interactions in ischemic
  preconditioning.
\newblock {\em The American journal of cardiology}, 80(3):124A--131A, 1997.

\bibitem{crea1990role}
Filippo Crea, Giuseppe Pupita, Alfredo~R Galassi, Hassan El-Tamimi, Juan~Carlos
  Kaski, Graham Davies, and Attilio Maseri.
\newblock Role of adenosine in pathogenesis of anginal pain.
\newblock {\em Circulation}, 81(1):164--172, 1990.

\bibitem{wijns2010guidelines}
William Wijns, Philippe Kolh, Nicolas Danchin, Carlo Di~Mario, Volkmar Falk,
  Thierry Folliguet, Scot Garg, Kurt Huber, Stefan James, Juhani Knuuti, et~al.
\newblock Guidelines on myocardial revascularization: the task force on
  myocardial revascularization of the european society of cardiology (esc) and
  the european association for cardio-thoracic surgery (eacts).
\newblock {\em European heart journal}, 31(20):2501--2555, 2010.

\bibitem{chamberlain1997coronary}
DA~Chamberlain, KAA Fox, RA~Henderson, DG~Julian, et~al.
\newblock Coronary angioplasty versus medical therapy for angina: the second
  randomised intervention treatment of angina (rita-2) trial.
\newblock {\em The Lancet}, 350(9076):461, 1997.

\bibitem{time2001trial}
Time Investigators et~al.
\newblock Trial of invasive versus medical therapy in elderly patients with
  chronic symptomatic coronary-artery disease (time): a randomised trial.
\newblock {\em The Lancet}, 358(9286):951--957, 2001.

\bibitem{weintraub2008effect}
William~S Weintraub, John~A Spertus, Paul Kolm, David~J Maron, Zefeng Zhang,
  Claudine Jurkovitz, Wei Zhang, Pamela~M Hartigan, Cheryl Lewis, Emir Veledar,
  et~al.
\newblock Effect of pci on quality of life in patients with stable coronary
  disease.
\newblock {\em New England Journal of Medicine}, 359(7):677--687, 2008.

\bibitem{erne2007effects}
Paul Erne, Andreas~W Schoenenberger, Dieter Burckhardt, Michel Zuber, Wolfgang
  Kiowski, Peter~T Buser, Paul Dubach, Therese~J Resink, and Matthias
  Pfisterer.
\newblock Effects of percutaneous coronary interventions in silent ischemia
  after myocardial infarction: the swissi ii randomized controlled trial.
\newblock {\em Jama}, 297(18):1985--1991, 2007.

\bibitem{boden2007optimal}
William~E Boden, Robert~A O'rourke, Koon~K Teo, Pamela~M Hartigan, David~J
  Maron, William~J Kostuk, Merril Knudtson, Marcin Dada, Paul Casperson,
  Crystal~L Harris, et~al.
\newblock Optimal medical therapy with or without pci for stable coronary
  disease.
\newblock {\em New England journal of medicine}, 356(15):1503--1516, 2007.

\bibitem{hueb2004medicine}
Whady Hueb, Paulo~R Soares, Bernard~J Gersh, Luiz~AM C{\'e}sar, Prot{\'a}sio~L
  Luz, Luiz~B Puig, Eul{\'o}gio~M Martinez, Sergio~A Oliveira, and Jos{\'e}~AF
  Ramires.
\newblock The medicine, angioplasty, or surgery study (mass-ii): a randomized,
  controlled clinical trial of three therapeutic strategies for multivessel
  coronary artery disease: one-year results.
\newblock {\em Journal of the American College of Cardiology},
  43(10):1743--1751, 2004.

\bibitem{stergiopoulos2012initial}
Kathleen Stergiopoulos and David~L Brown.
\newblock Initial coronary stent implantation with medical therapy vs medical
  therapy alone for stable coronary artery disease: meta-analysis of randomized
  controlled trials.
\newblock {\em Archives of internal medicine}, 172(4):312--319, 2012.

\bibitem{windecker2014authors}
Stephan Windecker, Philippe Kolh, F~Alfonso, JP~Collet, J~Cremer, V~Falk,
  G~Filippatos, C~Hamm, SJ~Head, P~J{\"u}ni, et~al.
\newblock Authors/task force members. 2014 esc/eacts guidelines on myocardial
  revascularization: the task force on myocardial revascularization of the
  european society of cardiology (esc) and the european association for
  cardio-thoracic surgery (eacts) developed with the special contribution of
  the european association of percutaneous cardiovascular interventions
  (eapci).
\newblock {\em Eur Heart J}, 35(37):2541--2619, 2014.

\bibitem{lamy2011medical}
Andre Lamy, Madhu Natarajan, and Salim Yusuf.
\newblock Medical treatment, pci, or cabg for coronary artery disease?, 2011.

\bibitem{blumenthal2000medical}
Roger~S Blumenthal, Gregory Cohn, and Steven~P Schulman.
\newblock Medical therapy versus coronary angioplasty in stable coronary artery
  disease: a critical review of the literature.
\newblock {\em Journal of the American College of Cardiology}, 36(3):668--673,
  2000.

\bibitem{pursnani2012percutaneous}
Seema Pursnani, Frederick Korley, Ravindra Gopaul, Pushkar Kanade, Newry
  Chandra, Richard~E Shaw, and Sripal Bangalore.
\newblock Percutaneous coronary intervention versus optimal medical therapy in
  stable coronary artery disease.
\newblock {\em Circulation: Cardiovascular Interventions}, pages
  CIRCINTERVENTIONS--112, 2012.

\bibitem{de2012fractional}
Bernard De~Bruyne, Nico~HJ Pijls, Bindu Kalesan, Emanuele Barbato, Pim~AL
  Tonino, Zsolt Piroth, Nikola Jagic, Sven M{\"o}bius-Winkler, Gilles Rioufol,
  Nils Witt, et~al.
\newblock Fractional flow reserve--guided pci versus medical therapy in stable
  coronary disease.
\newblock {\em New England Journal of Medicine}, 367(11):991--1001, 2012.

\bibitem{levine20112011}
Glenn~N Levine, Eric~R Bates, James~C Blankenship, Steven~R Bailey, John~A
  Bittl, Bojan Cercek, Charles~E Chambers, Stephen~G Ellis, Robert~A Guyton,
  Steven~M Hollenberg, et~al.
\newblock 2011 accf/aha/scai guideline for percutaneous coronary intervention:
  executive summary: a report of the american college of cardiology
  foundation/american heart association task force on practice guidelines and
  the society for cardiovascular angiography and interventions.
\newblock {\em Journal of the American College of Cardiology},
  58(24):2550--2583, 2011.

\bibitem{pijls1996measurement}
Nico~HJ Pijls, Bernard de~Bruyne, Kathinka Peels, Pepijn~H van~der Voort,
  Hans~JRM Bonnier, Jozef Bartunek, and Jacques~J Koolen.
\newblock Measurement of fractional flow reserve to assess the functional
  severity of coronary-artery stenoses.
\newblock {\em New England Journal of Medicine}, 334(26):1703--1708, 1996.

\bibitem{bruus2007theoretical}
Henrik Bruus.
\newblock {\em Theoretical microfluidics}.
\newblock Oxford university press Oxford, 2007.

\bibitem{klabunde2011cardiovascular}
Richard Klabunde.
\newblock {\em Cardiovascular physiology concepts}.
\newblock Lippincott Williams \& Wilkins, 2011.

\bibitem{schelbert2010anatomy}
Heinrich~R Schelbert.
\newblock Anatomy and physiology of coronary blood flow.
\newblock {\em Journal of nuclear cardiology}, 17(4):545--554, 2010.

\bibitem{muller1996integrated}
Judy~M Muller, Michael~J Davis, and William~M Chilian.
\newblock Integrated regulation of pressure and flow in the coronary
  microcirculation.
\newblock {\em Cardiovascular research}, 32(4):668--678, 1996.

\bibitem{tonino2009fractional}
Pim~AL Tonino, Bernard De~Bruyne, Nico~HJ Pijls, Uwe Siebert, Fumiaki Ikeno,
  Marcel vant Veer, Volker Klauss, Ganesh Manoharan, Thomas Engstr{\o}m,
  Keith~G Oldroyd, et~al.
\newblock Fractional flow reserve versus angiography for guiding percutaneous
  coronary intervention.
\newblock {\em N Engl j Med}, 2009(360):213--224, 2009.

\bibitem{pijls2007percutaneous}
Nico~HJ Pijls, Pepijn van Schaardenburgh, Ganesh Manoharan, Eric Boersma,
  Jan-Willem Bech, Marcel van’t Veer, Frits B{\"a}r, Jan Hoorntje, Jacques
  Koolen, William Wijns, et~al.
\newblock Percutaneous coronary intervention of functionally nonsignificant
  stenosis: 5-year follow-up of the defer study.
\newblock {\em Journal of the American College of Cardiology},
  49(21):2105--2111, 2007.

\bibitem{pijls2010fractional}
Nico~HJ Pijls, William~F Fearon, Pim~AL Tonino, Uwe Siebert, Fumiaki Ikeno,
  Bernhard Bornschein, Marcel van't Veer, Volker Klauss, Ganesh Manoharan,
  Thomas Engstr{\o}m, et~al.
\newblock Fractional flow reserve versus angiography for guiding percutaneous
  coronary intervention in patients with multivessel coronary artery disease:
  2-year follow-up of the fame (fractional flow reserve versus angiography for
  multivessel evaluation) study.
\newblock {\em Journal of the American College of Cardiology}, 56(3):177--184,
  2010.

\bibitem{lorell2000left}
Beverly~H Lorell and Blase~A Carabello.
\newblock Left ventricular hypertrophy.
\newblock {\em Circulation}, 102(4):470--479, 2000.

\bibitem{gunther1979determinants}
Stephen Gunther and William Grossman.
\newblock Determinants of ventricular function in pressure-overload hypertrophy
  in man.
\newblock {\em Circulation}, 59(4):679--688, 1979.

\bibitem{sugden1999signaling}
Peter~H Sugden.
\newblock Signaling in myocardial hypertrophy.
\newblock {\em Circulation Research}, 84(6):633--646, 1999.

\bibitem{borg1992holding}
TK~Borg and ML~Burgess.
\newblock Holding it all together: organization and function (s) of the
  extracellular matrix in the heart.
\newblock {\em Heart Failure}, 8:230--230, 1992.

\bibitem{kuppuswamy1997association}
Dhandapani Kuppuswamy, Charlene Kerr, Takahiro Narishige, Vijaykumar~S Kasi,
  Donald~R Menick, and George Cooper.
\newblock Association of tyrosine-phosphorylated c-src with the cytoskeleton of
  hypertrophying myocardium.
\newblock {\em Journal of Biological Chemistry}, 272(7):4500--4508, 1997.

\bibitem{terracio1991expression}
Louis Terracio, Kristofer Rubin, Donald Gullberg, ED~Balog, Wayne Carver, Ron
  Jyring, and Thomas~K Borg.
\newblock Expression of collagen binding integrins during cardiac development
  and hypertrophy.
\newblock {\em Circulation research}, 68(3):734--744, 1991.

\bibitem{tomanek2007angiogenesis}
Robert~J Tomanek and Eduard~I Dedkov.
\newblock Angiogenesis and arteriogenesis in cardiac hypertrophy.
\newblock {\em Modern concepts in angiogenesis}, pages 253--280, 2007.

\bibitem{duncker2008regulation}
Dirk~J Duncker and Robert~J Bache.
\newblock Regulation of coronary blood flow during exercise.
\newblock {\em Physiological reviews}, 88(3):1009--1086, 2008.

\bibitem{laughlin1985effects}
M~HAROLD Laughlin.
\newblock Effects of exercise training on coronary transport capacity.
\newblock {\em Journal of Applied Physiology}, 58(2):468--476, 1985.

\bibitem{laughlin2004physical}
M~Harold Laughlin.
\newblock Physical activity in prevention and treatment of coronary disease:
  the battle line is in exercise vascular cell biology.
\newblock {\em Medicine \& Science in Sports \& Exercise}, 36(3):352--362,
  2004.

\bibitem{treasure1993hypertension}
Charles~B Treasure, J~Larry Klein, Joseph~A Vita, Steven~V Manoukian, George~H
  Renwick, Andrew~P Selwyn, Peter Ganz, and R~Wayne Alexander.
\newblock Hypertension and left ventricular hypertrophy are associated with
  impaired endothelium-mediated relaxation in human coronary resistance
  vessels.
\newblock {\em Circulation}, 87(1):86--93, 1993.

\bibitem{pichard1981coronary}
August~D Pichard, Richard Gorlin, Harry Smith, John Ambrose, and Jose Meller.
\newblock Coronary flow studies in patients with left ventricular hypertrophy
  of the hypertensive type: evidence for an impaired coronary vascular reserve.
\newblock {\em The American journal of cardiology}, 47(3):547--554, 1981.

\bibitem{reed2017acute}
Grant~W Reed, Jeffrey~E Rossi, and Christopher~P Cannon.
\newblock Acute myocardial infarction.
\newblock {\em The Lancet}, 389(10065):197--210, 2017.

\bibitem{libby2013mechanisms}
Peter Libby.
\newblock Mechanisms of acute coronary syndromes and their implications for
  therapy.
\newblock {\em New England Journal of Medicine}, 368(21):2004--2013, 2013.

\bibitem{moreno2010vulnerable}
Pedro~R Moreno.
\newblock Vulnerable plaque: definition, diagnosis, and treatment.
\newblock {\em Cardiology clinics}, 28(1):1--30, 2010.

\bibitem{ghuran2001ischaemic}
AV~Ghuran and AJ~Camm.
\newblock Ischaemic heart disease presenting as arrhythmias.
\newblock {\em British medical bulletin}, 59(1):193--210, 2001.

\bibitem{chatterjee2014meta}
Saurav Chatterjee, Chirag Bavishi, Partha Sardar, Vikram Agarwal, Parasuram
  Krishnamoorthy, Tomasz Grodzicki, and Franz~H Messerli.
\newblock Meta-analysis of left ventricular hypertrophy and sustained
  arrhythmias.
\newblock {\em The American journal of cardiology}, 114(7):1049--1052, 2014.

\end{thebibliography}
\clearpage
\newpage
\begin{figure}[p!]\centering
\centering\captionsetup{format = hang}
\begin{measuredfigure}
\includegraphics[width=\textwidth]{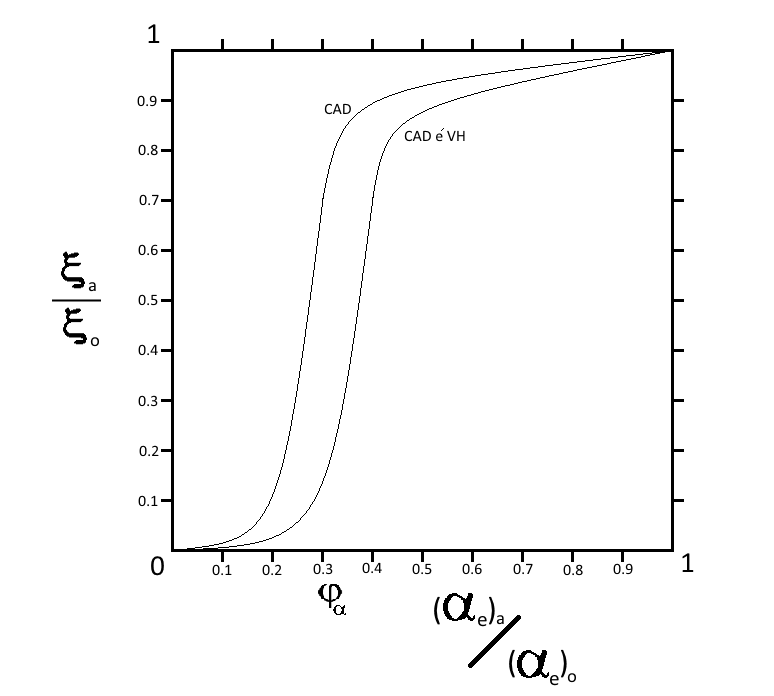}
\caption{A graphical representation of the relation between the ratio of the perfusion of myocardial tissue supplied by a stenotic epicardial coronary \(\xi_a\), to the perfusion in case of hypothetical absence of stenosis \(\xi_o\); and the ratio of the radius of said stenotic artery \((\alpha_e )_a\), to the radius in case of hypothetical absence of stenosis \((\alpha_e )_o\). Both ratios are presented by absolute numbers. In isolated CAD; the direct proportional relationship is represented by a sigmoid-shaped curve, wherein the perfusion of myocardial tissue supplied by said stenotic epicardial coronary \(\xi_a\) collapse relatively at a critical stenotic value \(\phi_\alpha\). Comorbid CAD and ventricular hypertrophy shift the curve to the right leading to an increase in the critical stenotic value \(\phi_\alpha\). }
\end{measuredfigure}
\end{figure}
\clearpage

\newpage
\begin{figure}[p!]\centering
\centering\captionsetup{format = hang}
\begin{measuredfigure}
\includegraphics[width=\textwidth]{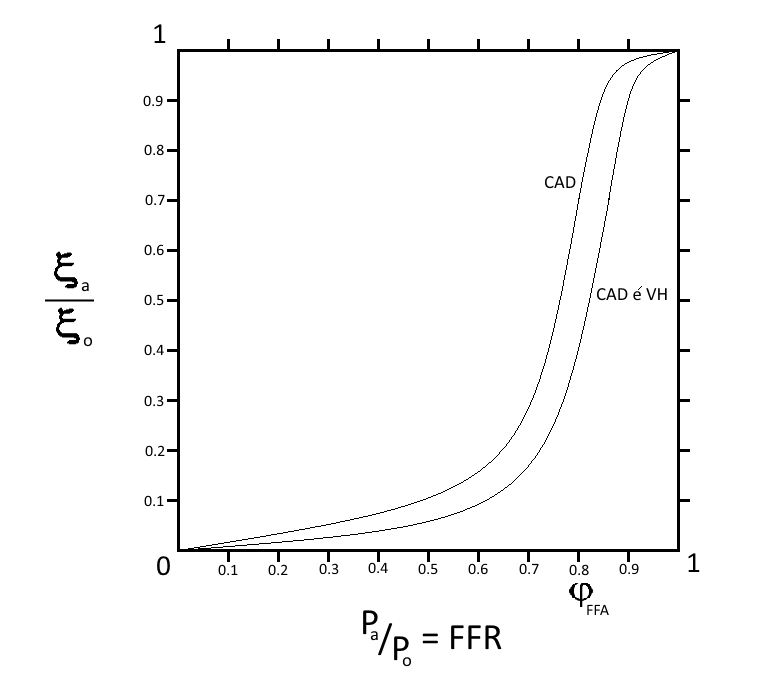}
\caption{A graphical representation of the relation between the ratio of the perfusion of myocardial tissue supplied by a stenotic epicardial coronary \(\xi_a\), to the perfusion in case of hypothetical absence of stenosis \(\xi_o\); and FFR which is the ratio of the pressure distal to the stenosis \(P_a\), to the pressure proximal to the stenosis \(P_o\). Both ratios are presented by absolute numbers. In isolated CAD; the direct proportional relationship is represented by a sigmoid-shaped curve, wherein the perfusion of myocardial tissue supplied by said stenotic epicardial coronary \(\xi_a\) collapse relatively at a critical stenotic value \(\phi_{FFR}\). Comorbid CAD and ventricular hypertrophy shift the curve to the right leading to an increase in the critical stenotic value \(\phi_{FFR}\).}
\end{measuredfigure}
\end{figure}
\clearpage
\end{document}